\documentclass[12pt,a4paper]{article}
\usepackage{amsmath,amssymb}
\usepackage{graphicx}
\usepackage{cite}

\allowdisplaybreaks[3]

\begin{document}

\begin{titlepage}

\begin{flushright}
NCTS-TH/1809
\end{flushright}

\vspace{5em}

\begin{center}
{\Large \textbf{Schwinger Effect in Inflaton-Driven Electric Field}}
\end{center}

\begin{center}
Hiroyuki \textsc{Kitamoto}\footnote{E-mail: kitamoto@cts.nthu.edu.tw}
\end{center}

\begin{center}
\textit{Physics Division, National Center for Theoretical Sciences}\\
\textit{National Tsing-Hua University, Hsinchu 30013, Taiwan}
\end{center}

\begin{abstract}
In a four dimensional inflation theory, a persistent electric field can be established by making the inflaton coupled to the gauge field like a dilaton.  
We investigate the pair production of scalar particles in the inflaton-driven electric field. 
In particular, we evaluate the induced current due to the pair production. 
The presence of the dilatonic coupling ensures the validity of the WKB approximation at the past and the future infinities, without tuning constant parameters. 
Thus, the semiclassical description is applicable in evaluating the induced current. 
Solving the field equations with the induced current, we evaluate the first-order backreaction to the electric field. 
It turns out that the electric field decreases with the cosmic expansion. 
The result indicates that the no-anisotropic hair theorem for inflation holds true regardless of whether the dilatonic coupling is present or not. 

\end{abstract}

\vspace{\fill}

Nov. 2018

\end{titlepage}

\section{Introduction}\label{Introduction}
\setcounter{equation}{0}

Concerning the primordial universe, we find no evidence of statistical anisotropy from the current status of cosmic microwave background observations \cite{Komatsu2013,Planck2015}. 
From a theoretical viewpoint, an anisotropic inflation can be obtained if a nonzero electromagnetic vector field exists. 
As a matter of fact, there is a following no-anisotropic hair theorem for inflation. 
In four dimensions, the gauge field respects the conformal symmetry as long as its kinetic term is canonical. 
The conformal symmetry ensures that the electromagnetic field decays as the inverse square of the scale factor of the Universe. 

S. Kanno, M. Watanabe and J. Soda showed that a persistent electric field can be obtained by introducing a dilatonic coupling between the inflaton and the gauge field in the action \cite{Soda2009}. 
We call it the inflaton-driven electric field.  
The dilatonic coupling breaks the conformal symmetry and so the no-anisotropic hair theorem discussed above is not applicable to this model. 
As pointed out in \cite{Martin2007}, it is natural to introduce such a noncanonical kinetic term of the gauge field from a perspective of the general supergravity action. 

The derivation of a persistent electric field in \cite{Soda2009} is based on the classical field equations. 
On the other hand, if a charged test field exists, a strong electric field leads to the pair production of particles \cite{Schwinger1951}. 
This is known as the Schwinger effect. 
The pair production induces the $U(1)$ current and the induced current screens the electric field, at least in Minkowski space. 
Therefore, it is reasonable to conjecture that if we take into account the Schwinger effect in the inflaton-driven electric field, the no-anisotropic hair theorem holds true also in the presence of the dilatonic coupling.  
In this paper, we verify the conjecture quantitatively by solving the field equations with the induced current. 

As a specific example of test fields, we consider a massive charged scalar field. 
We point out that the presence of the dilatonic coupling ensures the validity of the Wentzel--Kramers--Brillouin (WKB) approximation at the past and the future infinities, without tuning constant parameters.   
Based on this fact, we evaluate the induced current by using the semiclassical description. 

There are several studies which discussed the Schwinger effects in four dimensional de Sitter (dS) space \cite{Kobayashi2014,Hayashinaka2016-1,Hayashinaka2016-2,Hayashinaka2018,Soda2017}. 
In \cite{Kobayashi2014,Hayashinaka2016-1,Hayashinaka2016-2,Hayashinaka2018}, the physical electric field is fixed at a constant value in the absence of the dilatonic coupling. 
The definition of the physical electric field depends on whether the dilatonic coupling is present or not. 
That is, this paper investigates the Schwinger effect on a different background gauge field from \cite{Kobayashi2014,Hayashinaka2016-1,Hayashinaka2016-2,Hayashinaka2018}.  

In \cite{Soda2017}, the Schwinger effect has been investigated on the same background gauge field as this paper. 
The previous study discussed a weak electric field limit at the integrand level, where a momentum dependence is included. 
In evaluating integral quantities like the induced current, the parameter region can be interpreted as an early time. 
In contrast, this study discusses a late time behavior of the Schwinger effect. 

The organization of this paper is as follows. 
In Sec. \ref{Background}, we review the classical solution in the inflation theory with the dilatonic coupling. 
A persistent electric field is given as an attractor solution in this model. 
In Sec. \ref{Test}, we introduce a massive charged scalar field as a test field. 
We show that the WKB approximation is valid at the past and the future infinities. 
In Sec. \ref{IC}, we evaluate the induced current by using the semiclassical description. 
The evaluation is performed for a late time behavior. 
In Sec. \ref{BR}, we evaluate the first-order backreaction to the electric field by solving the field equations with the induced current.  
It turns out that the electric field decreases with the cosmic expansion. 
We conclude with discussions in Sec. \ref{Conclusion}. 

\section{Background fields}\label{Background}
\setcounter{equation}{0}

In order to obtain a nondecaying electromagnetic field during inflation, the action should include a term which breaks the conformal symmetry \cite{Soda2009,Martin2007,Turner1988,Ratra1991,Bamba2003,Bamba2007}. 
In this paper, we consider the inflation model where the inflaton $\varphi$ couples to the gauge field $A_\mu$ as  
\begin{align}
S_\text{bg}=\int\sqrt{-g}d^4x\ \left[\frac{M_\text{pl}^2}{2}R-\frac{1}{2}g^{\mu\nu}\partial_\mu\varphi\partial_\nu\varphi-V(\varphi)-\frac{1}{4}f^2(\varphi)g^{\mu\rho}g^{\nu\sigma}F_{\mu\nu}F_{\rho\sigma}\right], 
\label{model}\end{align}
\begin{align}
F_{\mu\nu}=\partial_\mu A_\nu-\partial_\nu A_\mu. 
\end{align}
Here $V$ is the inflaton potential dominating the cosmic energy density during the slow-roll inflation, and $f$ is the dilatonic coupling between the inflaton and the gauge field \cite{Soda2009,Martin2007}. 

The slow-roll condition for the inflaton potential approximates the background spacetime by dS space
\begin{align}
ds^2=-dt^2+a^2(t)d\textbf{x}^2,\hspace{1em}a(t)=e^{Ht},  
\end{align}
where $t$ is the cosmic time and $H$ is the Hubble parameter. 
The variation of $H$ and the anisotropic elements of the metric are suppressed by the slow-roll parameters. 

In this paper, we adopt the temporal gauge: 
\begin{align}
A_0=0. 
\label{temporal}\end{align}
Furthermore, we may solve the classical field equations along one spatial direction without loss of generality.  
The homogeneous background fields are written as 
\begin{align}
A_i=A(t)\delta_i^{\ 1},  
\label{spatial}\end{align}
\begin{align}
\varphi=\varphi(t),
\end{align}
where $i=1,2,3$. 
Then the equation of motion for $A$ is given by  
\begin{align}
\frac{d}{dt}\left(f^2a\frac{d}{dt}A\right)=0. 
\label{eom-A}\end{align}

As shown in \cite{Soda2009}, the time evolution of $f$ is determined as an attractor solution of the classical field equations. 
Specifically, the following ansatz is adopted: 
\begin{align}
f(\varphi)=\exp\left\{\frac{2c}{M_\text{pl}^2}\int d\varphi\ \frac{V}{\partial_\varphi V}\right\},
\label{ansatz}\end{align}
where $c$ is a constant parameter. 
As far as $c>1$, the classical solution approaches to 
\begin{align}
f=a^{-2}. 
\label{attractor}\end{align}
Here the overall coefficient is normalized. 

In the presence of the dilatonic coupling, the physical electric field is given by 
\begin{align}
E_\text{phys}=-fa^{-1}\frac{d}{dt}A. 
\end{align}
From (\ref{eom-A}) and (\ref{attractor}), we can confirm that the physical electric field approaches to a constant value 
\begin{align}
E_\text{phys}=E,\hspace{1em}E\text{: const.} 
\label{persistent}\end{align}
Here the constant value is related to the free parameter in (\ref{ansatz}) as $E=(\sqrt{3(c-1)\epsilon_V}/c) M_\text{pl}H$, $\epsilon_V\equiv (M_\text{pl}\partial_\varphi V/V)^2/2$. 
Integrating the equation, the gauge field is given by 
\begin{align}
A=-\frac{E}{3H}e^{3Ht}, 
\label{A}\end{align} 
where we kept the lowest order in $\epsilon_V$, 
i.e. we treated the background spacetime as dS space except that $E$ is proportional to $\sqrt{\epsilon_V}$.  
It should be noted that if the dilatonic coupling is absent as $f=1$, we cannot obtain a persistent electric field as a classical solution. 

In the above, we discussed the classical field equations. 
If a charged test field exists, a strong electric field leads to the pair production of particles, and the pair production induces the $U(1)$ current. 
Therefore, we need to consider the backreaction from the induced current to the electric field. 

The induced current has been evaluated in a different setting \cite{Kobayashi2014,Hayashinaka2016-1,Hayashinaka2016-2,Hayashinaka2018}. 
In the previous studies, the dilatonic coupling is absent and then the physical electric field is given by 
\begin{align}
\bar{E}_\text{phys}=-a^{-1}\frac{d}{dt}A. 
\end{align}
They discussed the case that the physical electric field is fixed at a constant value 
\begin{align}
\bar{E}_\text{phys}=E,\hspace{1em}E\text{: const.} 
\label{previous}\end{align}
In terms of the gauge field, the setting is expressed as 
\begin{align}
A=-\frac{E}{H}e^{Ht}. 
\label{A'}\end{align}
We emphasize that in \cite{Kobayashi2014,Hayashinaka2016-1,Hayashinaka2016-2,Hayashinaka2018}, a persistent electric field is considered as a setting rather than a classical solution. 
Considering the consistency with the background field equations, we adopt (\ref{persistent}) as a persistent electric field rather than (\ref{previous}).\footnote{
Note that the dimension of the spacetime is taken as $D=4$ in \cite{Kobayashi2014,Hayashinaka2016-1,Hayashinaka2016-2,Hayashinaka2018} and this paper. 
The induced current has been investigated also in the $D=2$ case \cite{Frob2014}, where a persistent electric field can be obtained without the dilatonic coupling.}

In the next section, we investigate the Klein--Gordon equation of a test field in the presence of the dilatonic coupling. 
The same Klein--Gordon equation has been investigated in \cite{Soda2017}. 
As a new finding, we show that the WKB approximation is valid not only at the past infinity but also at the future infinity.  
The validity originates in the presence of the dilatonic coupling and holds true for any values of constant parameters.  

\section{Test scalar field}\label{Test}
\setcounter{equation}{0}

As a test field, we introduce a massive charged scalar field $\phi$: 
\begin{align}
S_\text{test}=\int \sqrt{-g}d^4x\ \left[-g^{\mu\nu}(\partial_\mu+ieA_\mu)\phi^*(\partial_\nu-ieA_\nu)\phi-m^2\phi^*\phi\right]. 
\end{align}
In this paper, we consider quantum fluctuations of the test scalar field 
while we do not consider those of the other fields. 
Furthermore, we discuss quantum dynamics after the electric field becomes constant,  
i.e. we consider the case that the charge is sufficiently small 
so that the classical evolution of the electric field is much faster than the quantum one. 

In investigating the wave function, it is convenient to use the conformal time: 
\begin{align}
\tau=-\frac{1}{H}e^{-Ht},\hspace{1em}-\infty<\tau<0, 
\end{align}
and use the conformal transformation: 
\begin{align}
\tilde{\phi}(x)=a(\tau)\phi(x). 
\label{ct}\end{align}
From (\ref{temporal})--(\ref{spatial}) and (\ref{A}), the Klein--Gordon equation is given by 
\begin{align}
\left\{\frac{d^2}{d\tau^2}+\omega_\textbf{k}^2(\tau)\right\}\tilde{\phi}_\textbf{k}(x)=0, 
\label{KG1}\end{align}
\begin{align}
\omega_\textbf{k}^2(\tau)=\left\{k_1+\frac{eE}{3H(-H\tau)^3}\right\}^2+k_2^2+k_3^2+\frac{m^2-2H^2}{(-H\tau)^2}, 
\label{KG2}\end{align}
where $\tilde{\phi}_\textbf{k}$ is the wave function of $\tilde{\phi}$, 
and $\textbf{k}=(k_1,k_2,k_3)$ is the comoving momentum. 
We set $e>0$, $E>0$ in the subsequent discussion for simplicity, 
though parallel discussions can be applied to the other cases. 

In the Klein--Gordon equation, the background spacetime is approximated by dS space 
except that the electric field is nonzero due to a finite value of the slow-roll parameter. 
One purpose of this paper is to evaluate the time evolution of the induced current. 
As shown later, the time evolution is expressed as $a^\alpha,\ \alpha=\mathcal{O}(\epsilon_V^0)$. 
The approximation allows us to evaluate such a scaling, 
whose exponent is not suppressed by the slow-roll parameter. 

Although the Klein--Gordon equation (\ref{KG1})--(\ref{KG2}) is not exactly solvable, 
as far as the following discriminants are kept small 
\begin{align}
\omega_\textbf{k}^{-4}\left(\frac{d\omega_\textbf{k}}{d\tau}\right)^2\ll 1,\hspace{1em}
\left|\omega_\textbf{k}^{-3}\frac{d^2\omega_\textbf{k}}{d\tau^2}\right|\ll 1, 
\end{align}
we may adopt the WKB approximation: 
\begin{align}
\tilde{\phi}_\textbf{k}(x)\simeq \frac{1}{\sqrt{2\omega_\textbf{k}(\tau)}}\exp\left\{-i\int^\tau d\tau'\ \omega_\textbf{k}(\tau')\right\}e^{+i\textbf{k}\cdot\textbf{x}}. 
\end{align}

At the past infinity $\tau\to-\infty$, $\omega_\textbf{k}$ is constant 
\begin{align}
\omega_\textbf{k}\simeq |\textbf{k}|, 
\label{past}\end{align}
and then the validity of the WKB approximation is trivial 
\begin{align}
\omega_\textbf{k}^{-4}\left(\frac{d\omega_\textbf{k}}{d\tau}\right)^2\simeq 0,\hspace{1em}
\omega_\textbf{k}^{-3}\frac{d^2\omega_\textbf{k}}{d\tau^2}\simeq 0. 
\end{align}

At the future infinity $\tau\to 0$, the electric field term of $\omega_\textbf{k}$ becomes dominant as   
\begin{align}
\omega_\textbf{k}\simeq \frac{eE}{3H(-H\tau)^3}, 
\label{future}\end{align}
and then the discriminants approach to zero as 
\begin{align}
&\omega_\textbf{k}^{-4}\left(\frac{d\omega_\textbf{k}}{d\tau}\right)^2\simeq 9\left\{\frac{eE}{3H^2(-H\tau)^2}\right\}^{-2}, \notag\\
&\omega_\textbf{k}^{-3}\frac{d^2\omega_\textbf{k}}{d\tau^2}\simeq 12\left\{\frac{eE}{3H^2(-H\tau)^2}\right\}^{-2}. 
\label{decay}\end{align}
The decay of the discriminants shows that the WKB approximation is valid at the future infinity for any values of $m^2/H^2$ and $eE/H^2$ (except for $eE/H^2=0$). 

It should be noted that the decay of the discriminants is due to the presence of the dilatonic coupling. 
For comparison, let us consider the case that the physical electric field is fixed in the absence of the dilatonic coupling as in (\ref{previous}). 
From (\ref{temporal})--(\ref{spatial}) and (\ref{A'}), the Klein--Gordon equation is given by 
\begin{align}
\left\{\frac{d^2}{d\tau^2}+\bar{\omega}_\textbf{k}^2(\tau)\right\}\tilde{\phi}_\textbf{k}(x)=0, 
\label{KG1'}\end{align}
\begin{align}
\bar{\omega}_\textbf{k}^2(\tau)=\left\{k_1+\frac{eE}{H(-H\tau)}\right\}^2+k_2^2+k_3^2+\frac{m^2-2H^2}{(-H\tau)^2}. 
\label{KG2'}\end{align}
The validity of the WKB approximation is trivial at the past infinity as well as in (\ref{KG1})--(\ref{KG2}). 

At the future infinity, $\bar{\omega}_\textbf{k}$ behaves as 
\begin{align}
\bar{\omega}_\textbf{k}\simeq\frac{1}{-H\tau}\left\{\left(\frac{eE}{H^2}\right)^2+\frac{m^2}{H^2}-2\right\}^\frac{1}{2}, 
\label{future'}\end{align}
and then the discriminants approach to the constant values 
\begin{align}
&\bar{\omega}_\textbf{k}^{-4}\left(\frac{d\bar{\omega}_\textbf{k}}{d\tau}\right)^2\simeq \left\{\left(\frac{eE}{H^2}\right)^2+\frac{m^2}{H^2}-2\right\}^{-1}, \notag\\
&\bar{\omega}_\textbf{k}^{-3}\frac{d^2\bar{\omega}_\textbf{k}}{d\tau^2}\simeq 2\left\{\left(\frac{eE}{H^2}\right)^2+\frac{m^2}{H^2}-2\right\}^{-1}. 
\end{align}
It should be noted that there exists the following constant parameter region where the WKB approximation is invalid: 
\begin{align}
\left(\frac{eE}{H^2}\right)^2\lesssim 1,\hspace{1em}
\frac{m^2}{H^2}\lesssim 1. 
\label{non-WKB}\end{align}
The induced current has been evaluated in the non-semiclassical region \cite{Frob2014,Kobayashi2014,Hayashinaka2016-1,Hayashinaka2016-2,Hayashinaka2018}.
In the $D=4$ case, it is claimed that in a weak electric field limit of (\ref{non-WKB}), the induced current 
antiscreens the electric field \cite{Kobayashi2014,Hayashinaka2016-1,Hayashinaka2016-2,Hayashinaka2018}. 

Comparing (\ref{KG2}) with (\ref{KG2'}), it turns out that the dilatonic coupling gives an additional scaling to the electric field term as 
\begin{align}
\frac{eE}{H^2}\ \to\ \frac{eE}{H^2(-H\tau)^2}. 
\end{align}
The additional scaling causes the decay of the discriminants in (\ref{decay}). 
We can conclude that if the dilatonic coupling is present, the non-semiclassical region becomes extinct at the future infinity without tuning the values of $m^2/H^2$ and $eE/H^2$.  
Thus, in the next section, we evaluate the induced current by using the semiclassical description. 

\section{Induced current due to pair production}\label{IC}
\setcounter{equation}{0}

As seen in (\ref{past}) and (\ref{future}), the frequency shows different behaviors at the past and the future infinities. 
This fact shows that the inflaton-driven electric field leads to the pair production of particles.   
In this section, we evaluate the induced current due to the pair production. 

If the WKB approximation is valid at the past and the future infinities, the produced particle number can be evaluated by considering the turning point of the frequency in the complex time plane. 
Specifically, the produced particle number is expressed as 
\begin{align}
n_\textbf{k}=\exp\left\{4\ \text{Im}\int^{\tau_*} d\tau'\ \omega_\textbf{k}(\tau')\right\}, 
\label{sc1}\end{align}
where $\tau_*$ is defined as the frequency vanishes at the complex time 
\begin{align}
\omega_\textbf{k}(\tau_*)\equiv 0. 
\label{sc2}\end{align}
Please refer to \cite{Pokrovskii1961} for the derivation of the expression. 
The evaluation method is also adopted to investigate the pair production in global dS space \cite{Mottola2013}. 

The $U(1)$ current of the test scalar field is given by 
\begin{align}
j_\mu=-ie\left[\langle\phi^*(\partial_\mu-ieA_\mu)\phi\rangle-\langle\phi(\partial_\mu+ieA_\mu)\phi^*\rangle\right]. 
\end{align}
For convenience, we consider the rescaled current $\tilde{j}_\mu$ of the canonically normalized scalar field (\ref{ct}): 
\begin{align}
\tilde{j}_\mu=a^2j_{\mu}. 
\end{align}
For the background field configuration (\ref{temporal})--(\ref{spatial}), the current is written as 
\begin{align}
\tilde{j}_0=0,\hspace{1em}\tilde{j}_i=\tilde{j}(\tau)\delta_i^{\ 1}. 
\end{align}

Using the semiclassical picture (\ref{sc1})--(\ref{sc2}), we can express the induced current as 
\begin{align}
\tilde{j}(\tau)=2e\int\frac{d^3k}{(2\pi)^3}\ v_\textbf{k}(\tau)n_\textbf{k},
\label{j}\end{align}
\begin{align}
v_\textbf{k}(\tau)=\left\{k_1+\frac{eE}{3H(-H\tau)^3}\right\}/\omega_\textbf{k}(\tau). 
\end{align}
Note that only the dominant term is shown in (\ref{j}). 
The general from of the current is  
\begin{align}
\tilde{j}(\tau)=2e\int\frac{d^3k}{(2\pi)^3}\ v_\textbf{k}(\tau)
\left[\frac{1}{2}
+\text{Re}\left\{\alpha_\textbf{k}\beta_\textbf{k}^* e^{-2i\int^\tau d\tau'\ \omega(\tau')}\right\}
+|\beta_\textbf{k}|^2\right],
\end{align}
where $\alpha_\textbf{k},\beta_\textbf{k}$ are the Bogoliubov coefficients 
and $|\beta_\textbf{k}|^2=n_\textbf{k}$. 
Due to the presence of the oscillating factor $e^{-2i\int^\tau d\tau'\ \omega(\tau')}$, 
after the momentum integral, 
the second term gives a relatively small contribution compared with the other two terms. 
The first term is the contribution from the vacuum 
and the momentum integral of it is ultraviolet divergent.   
After canceling the ultraviolet divergence by a counter term \cite{Herman1996}, 
we obtain a finite contribution which is proportional to 
\begin{align}
a^2e^2\nabla_\alpha F^\alpha_{\ \mu}\propto a^3eH^3\left(\frac{eE}{H^2}a^2\right)^1, 
\end{align}
where $\nabla_a$ is the covariant derivative for spacetime. 
As shown later, the contribution from the third term is proportional to $a^3eH^3\left(\frac{eE}{H^2}a^2\right)^2$. 
So, the late time behavior of the induced current is well described by (\ref{j}). 

In evaluating the produced particle number, we divide (\ref{KG2}) into the two parts as 
\begin{align}
\omega_\textbf{k}^2(\tau)=A^2_\textbf{k}(\tau)+B_\textbf{k}(\tau), 
\end{align}
\begin{align}
A_\textbf{k}(\tau)&\equiv k_1+\frac{eE}{3H(-H\tau)^3}, \notag\\
B_\textbf{k}(\tau)&\equiv k_2^2+k_3^2+\frac{m^2-2H^2}{(-H\tau)^2}. 
\end{align}
We consider the long time scale where the $B_\textbf{k}$ part can be treated as a perturbation from the $A_\textbf{k}^2$ part. 
For the turning point in the complex time plane, the $A_\textbf{k}^2$ part dominantly determines $\text{Re}\ \tau_*$, while the $B_\textbf{k}$ part determines $\text{Im}\ \tau_*$. 
That is, $\tau_*$ is approximated as 
\begin{align}
\tau_*\simeq -\frac{1}{H}\left(\frac{-eE}{3k_1H}\right)^\frac{1}{3}-i\epsilon, 
\label{tp}\end{align}
where $\epsilon$ is $-\text{Im}\ \tau_*$ and is taken positive for a convergence. 
The explicit form of $\epsilon$ is not necessary for the evaluation of $n_\textbf{k}$ and so we do not show it here. 

In evaluating the imaginary part of the integral in (\ref{sc1}), the first-order expansion of the frequency is crucial 
\begin{align}
\omega_\textbf{k}(\tau)&=A_\textbf{k}(\tau)+\frac{1}{2}\frac{B_\textbf{k}(\tau)}{A_\textbf{k}(\tau)}+\cdots. 
\label{expansion}\end{align}
For the $B_\textbf{k}/A_\textbf{k}$ term, we may approximate its denominator and numerator as 
\begin{align}
A_\textbf{k}(\tau)\simeq (\tau-\text{Re}\ \tau_*)\cdot \frac{d}{d\tau}A_\textbf{k}(\tau)\big|_{\tau=\text{Re}\ \tau_*}, 
\label{denominator}\end{align}
\begin{align}
B_\textbf{k}(\tau)&\simeq B_\textbf{k}(\text{Re}\ \tau_*). 
\label{numerator}\end{align}
From (\ref{tp})--(\ref{numerator}), the produced particle number is evaluated as 
\begin{align}
n_\textbf{k}&\simeq \exp\left\{2\ \text{Im}\int^{\tau_*} d\tau'\ \frac{B_\textbf{k}(\text{Re}\ \tau_*)}{(\tau'-\text{Re}\ \tau_*)\cdot\frac{d}{d\tau'}A_\textbf{k}(\tau')\big|_{\tau'=\text{Re}\ \tau_*}}\right\} \notag\\
&=\exp\left\{-\frac{\pi B_\textbf{k}(\text{Re}\ \tau_*)}{\frac{d}{d\tau}A_\textbf{k}(\tau)\big|_{\tau=\text{Re}\ \tau_*}}\right\} \notag\\
&=\exp\left[-\frac{\pi}{3}\left\{\frac{k_2^2+k_3^2}{k_1^2\left(\frac{k_1^2eE}{3H^4}\right)^{-\frac{1}{3}}}+\frac{m^2-2H^2}{H^2\left(\frac{k_1^2eE}{3H^4}\right)^\frac{1}{3}}\right\}\right]. 
\label{n}\end{align}
In the second line, $\text{Im}\ \tau_*$ gives a nonzero contribution as $\log(-i\epsilon)=-i\pi/2$. 

After evaluating the produced particle number, we may neglect the $B_\textbf{k}$ part of $\omega_\textbf{k}$ and then the velocity is approximately unity 
\begin{align}
v_\textbf{k}\simeq 1. 
\label{v}\end{align}
Substituting (\ref{n}) and (\ref{v}) to (\ref{j}), the induced current is written as 
\begin{align}
\tilde{j}(\tau)&\simeq 2e\int \frac{d^3k}{(2\pi)^3}\ \exp\left[-\frac{\pi}{3}\left\{\frac{k_2^2+k_3^2}{k_1^2\left(\frac{k_1^2eE}{3H^4}\right)^{-\frac{1}{3}}}+\frac{m^2-2H^2}{H^2\left(\frac{k_1^2eE}{3H^4}\right)^\frac{1}{3}}\right\}\right] \notag\\
&=\frac{2e}{(2\pi)^3}\int dk_1\ 3k_1^2\left(\frac{k_1^2eE}{3H^4}\right)^{-\frac{1}{3}}\exp\left\{-\frac{\pi}{3}\frac{m^2-2H^2}{H^2\left(\frac{k_1^2eE}{3H^4}\right)^\frac{1}{3}}\right\}. 
\end{align}
In the second line, we performed the Gaussian integrals with respect to $k_2$ and $k_3$. 

For the remaining $k_1$ integral, it should be recalled that the pair production occurs after $A_\textbf{k}(\tau)$ exceeds zero\footnote{
Strictly speaking, there is the additional condition for the initial time $\tau_0$: $A(\tau_0)<0$. 
The initial condition adds a constant term in the induced current. 
We consider the long time scale: $a(\tau)\gg a(\tau_0)$, where the constant term is negligible. }
\begin{align}
A_\textbf{k}(\tau)=k_1 + \frac{eE}{3H(-H\tau)^3}> 0. 
\label{cutoff}\end{align}
The time dependent cutoff is consistent with the conservation of the current: 
\begin{align}
g^{\mu\nu}\nabla_\mu j_\nu=0
\hspace{1em}\Leftrightarrow\hspace{1em}
\eta^{\mu\nu}\partial_\mu \tilde{j}_\nu=0. 
\end{align}
That is because $\partial_0$ does not act on the cutoff inside $\tilde{j}_i$.  
After performing the $k_1$ integral with (\ref{cutoff}), we obtain 
\begin{align}
\tilde{j}(\tau)\simeq \frac{e^3E^2}{4\pi^3}\frac{a^7(\tau)}{7H}\exp\left\{-\pi\frac{m^2-2H^2}{eEa^2(\tau)}\right\}. 
\label{j-result0}\end{align}
Here we focus on the behavior at the late time: 
\begin{align}
\frac{eE}{H^2}a^2(\tau)\gg 1. 
\label{late1}\end{align}
We clarify the difference between this study and the previous study \cite{Soda2017} which investigated the Schwinger effect on the same background gauge field. 
The previous study discussed the weak electric field limit: $k_1^2eE/H^4\ll 1$ at the integrand level. 
In evaluating integral quantities like the induced current, the parameter region can be interpreted as the early time: $\frac{eE}{H^2}a^2(\tau)\ll 1$. 
In contrast, this study discussed the late time behavior of the induced current as seen in (\ref{tp})--(\ref{late1}). 

Furthermore, after enough time has passed 
\begin{align} 
\frac{|m^2-2H^2|}{eEa^2(\tau)}\ll 1, 
\label{late2}\end{align}
we may approximate the exponential factor in (\ref{j-result0}) by unity as 
\begin{align}
\tilde{j}(\tau)\simeq \frac{e^3E^2}{4\pi^3}\frac{a^7(\tau)}{7H}.  
\label{j-result1}\end{align}
It should be noted that the approximation at late time holds true regardless of the sign of $(m^2-2H^2)$. 
We also emphasize that the decay of the exponent is due to the presence of the dilatonic coupling. 
As seen in \cite{Kobayashi2014,Hayashinaka2016-1,Hayashinaka2016-2,Hayashinaka2018}, the exponent is time independent if the physical electric field is fixed in the absence of the dilatonic coupling. 

Since the dynamics of the electric field $E_\text{phys}$ is not independent with that of the dilatonic factor $f$, the presence of the induced current does not always lead to the decay of the electric field. 
For example, if the contribution from the induced current were compensated completely by the deformation of the dilatonic factor, the electric field should be persistent. 
In the next section, we solve the simultaneous field equations with the induced current (\ref{j-result1}) to verify the no-anisotropic hair theorem. 

\section{Backreaction from induced current}\label{BR}
\setcounter{equation}{0}

Including the induced current, the field equations are given by 
\begin{align}
V=3M_\text{pl}^2H^2, 
\label{eq1}\end{align}
\begin{align}
3H\frac{d}{dt}\varphi+\partial_\varphi V-f^{-1}\partial_\varphi f \cdot E_\text{phys}^2=0, 
\label{eq2}\end{align}
\begin{align}
\frac{d}{dt}(f a^2 E_\text{phys})+a^{-1}\tilde{j}=0, 
\label{eq3}\end{align}
where the slow-roll condition is imposed. 
By adopting the ansatz (\ref{ansatz}), we can express the derivatives of $f$ as  
\begin{align}
\partial_\varphi f=\frac{2c}{\sqrt{2\epsilon_V}M_\text{pl}}f,\hspace{1em}\frac{d}{dt}f=\frac{d}{dt}\varphi\cdot\partial_\varphi f. 
\label{ansatz'}\end{align}
We repeatedly make use of the relations in solving the field equations. 

The equation (\ref{eq3}) is integrated as 
\begin{align}
E_\text{phys}=f^{-1}a^{-2}\left(E-\int^t_{t_0}dt'\ a^{-1}(t')\tilde{j}(t')\right). 
\label{eq3'}\end{align}
As discussed later, we normalize the overall coefficient of $f$ and so the integration constant is equal to the initial value of the electric field $E$. 

From (\ref{eq1}), (\ref{eq2}) and (\ref{eq3'}), we obtain 
\begin{align}
\left(\frac{d}{dt}\varphi+\sqrt{2\epsilon_V}M_\text{pl}H\right)f^2a^4
&=\frac{2c}{3\sqrt{2\epsilon_V}M_\text{pl}H}\left(E-\int^t_{t_0}dt'\ a^{-1}(t')\tilde{j}(t')\right)^2 \notag\\
&\simeq \frac{2c}{3\sqrt{2\epsilon_V}M_\text{pl}H}\left(E^2-2E\int^t_{t_0}dt'\ a^{-1}(t')\tilde{j}(t')\right). 
\label{key0}\end{align}
In the second line, we evaluated the right side in the first-order approximation. 
That is because the induced current is evaluated on the classical background in this paper. 
Integrating (\ref{key0}), we obtain 
\begin{align}
f^2a^4=a^{-4(c-1)}\int dt\ a^{4(c-1)}\frac{4c^2}{3\epsilon_VM_\text{pl}^2H}\left(E^2-2E\int^t_{t_0}dt'\ a^{-1}(t')\tilde{j}(t')\right), 
\label{key1}\end{align}
where the first integral means an indefinite integral. 

If any charged test field does not exist, 
\begin{align}
f^2a^4&=a^{-4(c-1)}\int dt\ a^{4(c-1)}\frac{4c^2}{3\epsilon_VM_\text{pl}^2H}E^2 \notag\\
&=\frac{c^2E^2}{3\epsilon_V(c-1)M_\text{pl}^2H^2}+qa^{-4(c-1)},  
\label{key'}\end{align}
where $q$ is an integration constant. 
It should be recalled that we keep the lowest order in $\epsilon_V$. 
We consider the $c>1$ case where the contribution from the integration constant decays to zero.\footnote{
In the $c<1$ case, $f\propto a^{-2c}$ and then $E_\text{phys}\propto a^{2(c-1)}$. We cannot obtain a persistent electric field.} 
Normalizing the overall coefficient of $f$, we can identify the value of $E$ as
\begin{align}
E=\frac{\sqrt{3\epsilon_V(c-1)}}{c}M_\text{pl}H. 
\end{align} 
From (\ref{key'}) and (\ref{eq3'}), the dilatonic factor and the electric field are given by 
\begin{align}
f=a^{-2},\hspace{1em}E_\text{phys}=E. 
\end{align}

Let us go back to the case where a charged test field exists. 
Substituting the explicit form (\ref{j-result1}), the contribution from the induced current is evaluated as 
\begin{align}
\int^t_{t_0}dt'\ a^{-1}(t')\tilde{j}(t')\simeq \frac{e^3E^2}{4\pi^3}\frac{a^6}{42H^2}, 
\label{j-result2}\end{align}
where we consider the long time scale: $a(t)\gg a(t_0)$. 
From (\ref{key1}) and (\ref{eq3'}) including (\ref{j-result2}), the dilatonic factor and the electric field are given by 
\begin{align}
f=a^{-2}\left\{1-\frac{1}{1+\frac{3}{2}\frac{1}{c-1}}\cdot\frac{e^3E}{4\pi^3}\frac{a^6}{42H^2}\right\}, 
\label{dilatonic}\end{align}
\begin{align}
E_\text{phys}=E\left\{1-\frac{\frac{3}{2}\frac{1}{c-1}}{1+\frac{3}{2}\frac{1}{c-1}}\cdot\frac{e^3E}{4\pi^3}\frac{a^6}{42H^2}\right\}. 
\label{electric}\end{align}
The induced current provides additional scalings not only to the dilatonic factor but also to the electric field.\footnote{
If the contribution from the induced current were given as $\int^t_{t_0}dt'\ a^{-1}(t')\tilde{j}(t')\propto \log a$, 
such a slow evolution should be compensated completely by the deformation of the dilatonic factor and so the electric field should be persistent.}
We emphasize that the coefficients of the additional terms are negative in the discussed parameter region $c>1$. 
Therefore, we can conclude that the electric field decreases with the cosmic expansion due to the induced current. 
The result indicates that as far as a charged test scalar field exists, the no-anisotropic hair theorem holds true also in the inflation theory with the dilatonic coupling. 

It should be recalled that in this paper, the backreaction from the induced current is evaluated in the first-order approximation. 
Specifically, the investigation is valid for describing the dynamics at the initial stage: 
\begin{align}
e\left(\frac{E}{H^2}\right)^\frac{1}{3}a^2(t)\ll 1. 
\label{first}\end{align}
From the observed scalar amplitude $A_s\sim 2\times 10^{-9}$, $E/H^2$ is estimated as $E/H^2\sim 4\times 10^3\times\sqrt{(c-1)/c}$. 
Thus, there exists the parameter region satisfying (\ref{late1}) and (\ref{first}) simultaneously unless $(c-1)$ is fine-tuned to a tiny value.\footnote{
If we do not take into account the induced current, $(c-1)$ is constrained by the observational limit of the statistical anisotropy $g_*$ 
as $c-1\lesssim 10^{-7}\times(g_*/10^{-2})\times(N/60)^{-2}$ where $N$ is the e-folding number \cite{Soda2010}. 
Since the induced current screens the electric field, we do not fine-tune the value of $(c-1)$ here.}
It is a future subject to investigate the whole time evolutions of the dilatonic factor and the electric field. 
For the investigation, we need to evaluate the induced current on a general background. 

\section{Conclusion}\label{Conclusion}
\setcounter{equation}{0}

In the inflation theory with a dilatonic coupling between the inflaton and the gauge field, a persistent electric field is given as an attractor solution of the classical field equations \cite{Soda2009}. 
In other words, in this model, the no-anisotropic hair theorem for inflation does not hold true at the classical level. 
In order to verify the no-anisotropic hair theorem at the quantum level, we investigated the pair production of scalar particles in the inflaton-driven electric field. 
Specifically, we evaluated the induced current due to the pair production and evaluated the backreaction from the induced current to the electric field. 

We found that the presence of the dilatonic coupling ensures the validity of the WKB approximation not only at the past infinity but also at the future infinity, without tuning the values of $m^2/H^2$ and $eE/H^2$. 
Based on this fact, we evaluated the produced particle number by considering the turning point of the frequency in the complex time plane. 
Furthermore, we evaluated the induced current by using the semiclassical description. 
In contrast to the previous study \cite{Soda2017}, this study evaluated the behavior of the induced current at the late time: $a^2\gg H^2/(eE),\ |m^2-2H^2|/(eE)$.  

We evaluated the first-order backreaction to the background by solving the field equations with the induced current. 
The investigation is valid for describing the dynamics at the initial stage: $a^2\ll (H^2/E)^\frac{1}{3}/e$. 
Since the contribution from the induced current evolves rapidly as $\int^t_{t_0}dt'\ a^{-1}(t')\tilde{j}(t')\propto a^6$, 
the rapid evolution is not compensated completely by the deformation of the dilatonic factor and it screens the electric field with the cosmic expansion. 
The result indicates that as far as a charged scalar field exists, the no-anisotropic hair theorem holds true also in the inflation theory with the dilatonic coupling.  

It is a future subject to investigate the whole time evolutions of the dilatonic factor and the electric field. 
Such a nonperturbative investigation is necessary to prove the no-anisotropic hair theorem completely. 
For the investigation, we need to evaluate the induced current on a general background. 
We emphasize that as far as the background satisfies the validity conditions of the WKB approximation, the semiclassical description can simplify the evaluation of the induced current as seen in this paper. 

The fermionic pair production in the inflaton-driven electric field is another open problem. 
In contrast to scalar fields, Dirac fields do not have supercurvature modes for any value of the mass. 
On the other hand, if the dilatonic coupling is present, the electric field term in the Klein--Gordon equation is dominant compared with the mass term at late time. 
Thus, we conjecture that there is no significant difference between the pair production of scalar particles and that of fermions. 

As an aside, we mention the studies of non-Abelian gauge fields during inflation. 
By introducing a $(\epsilon^{\mu\nu\rho\sigma}F^a_{\mu\nu}F^a_{\rho\sigma})^2$ term in the Yang--Mills action \cite{Maleknejad2011-1,Maleknejad2011-2}, 
or an axion $\chi$ interacting with the gauge field through a $\chi(\epsilon^{\mu\nu\rho\sigma}F^a_{\mu\nu}F^a_{\rho\sigma})$ term \cite{Adshead2012}, 
the classical field equations lead to the gauge field which is linearly proportional to the scale factor as $A_i^a\propto a\delta_i^a$. 
Here $F^a_{\mu\nu}$ is the field strength of the $SU(2)$ gauge field $A^a_\mu$, $a=1,2,3$ and $\epsilon^{\mu\nu\rho\sigma}$ is the completely antisymmetric tensor. 
The scaling of the gauge field is the same as that in \cite{Kobayashi2014,Hayashinaka2016-1,Hayashinaka2016-2,Hayashinaka2018}, and different from that in \cite{Soda2017} and this paper.  
The Schwinger effect in the $SU(2)$ gauge field was investigated recently \cite{Komatsu2018}. 
Unlike in the $U(1)$ case, the nonzero gauge field and the induced current respect the isotropy. 
Furthermore, it was claimed that the $SU(2)$ current screens the electromagnetic field for any values of constant parameters. 

\section*{Acknowledgment}

This work is supported by the National Center of Theoretical Sciences (NCTS). 
We thank Chong-Sun Chu, Hiroyuki Ishida, Yoji Koyama and Jiro Soda for discussions. 
We thank in particular Jiro Soda for reading the manuscript and for valuable comments. 
His comments helped us to improve our understanding of the backreaction from the induced current to the electric field. 


\end{document}